\documentclass[useAMS,usenatbib]{mn2e}
\usepackage{color}
\usepackage{tabularx}
\usepackage{multirow}
\usepackage{amsmath}
\usepackage[para,online,flushleft]{threeparttable}
\usepackage[dvipdfmx]{graphicx}

\usepackage{color}

\title[Two-component jet model for GRB 221009A]{
Two-component jet model for multiwavelength afterglow emission of
the extremely energetic burst GRB~221009A
}
\author[Y. Sato et al.]
{Yuri~Sato$^{1}$\thanks{E-mail: yuris@phys.aoyama.ac.jp (YS)},
Kohta~Murase$^{2,3,4,5,6}$,
Yutaka~Ohira$^{7}$
and Ryo~Yamazaki$^{1,8}$\\
$^{1}$Department of Physical Sciences, Aoyama Gakuin University, 5-10-1 Fuchinobe, Sagamihara 252-5258, Japan\\
$^{2}$Department of Physics, Pennsylvania State University, University Park, Pennsylvania 16802, USA\\
$^{3}$Department of Astronomy \& Astrophysics, Pennsylvania State University, University Park, Pennsylvania 16802, USA \\
$^{4}$Center for Multimessenger Astrophysics, Pennsylvania State University, University Park, Pennsylvania 16802, USA\\
$^{5}$School of Natural Sciences, Institute for Advanced Study, Princeton, NJ 08540, USA \\
$^{6}$Center for Gravitational Physics and Quantum Information, Yukawa Institute for Theoretical Physics, Kyoto University, Kyoto,\\ Kyoto 606-8502, Japan\\
$^{7}$Department of Earth and Planetary Science, The University of Tokyo, 7-3-1 Hongo, Bunkyo-ku, Tokyo 113-0033, Japan,\\
$^{8}$Institute of Laser Engineering, Osaka University, 2-6 Yamadaoka, Suita, Osaka 565-0871, Japan
}
\begin{document} 
\pagerange{\pageref{firstpage}--\pageref{lastpage}} \pubyear{2023}
\maketitle
\label{firstpage}

\begin{abstract}
Recently gamma-ray bursts (GRBs) have been detected at very high-energy (VHE) gamma-rays by imaging atmospheric Cherenkov telescopes, and a two-component jet model has often been invoked to explain multiwavelength data. In this work, multiwavelength afterglow emission from an extremely bright GRB, GRB~221009A, is examined. 
The isotropic-equivalent gamma-ray energy of this event
is among the largest, which suggests that similarly to previous VHE GRBs, the jet opening angle is so small that the collimation-corrected gamma-ray energy is nominal. 
Afterglow emission from such a narrow jet decays too rapidly, especially if the jet propagates into uniform circumburst material. In the two-component jet model, another wide jet component with a smaller Lorentz factor dominates late-time afterglow emission, and we show that multiwavelength data of GRB 221009A can be explained by narrow and wide jets with opening angles similar to those employed for other VHE GRBs. 
We also discuss how model degeneracies can be disentangled with observations. 
\end{abstract}
 


\section{Introduction}
In recent years, very-high-energy (VHE) gamma-ray photons from some gamma-ray bursts (GRBs) have been detected ~\citep{MAGIC2019a,MAGIC2019b,HESS2019,Blanch2020,HESS2021}.
The observed VHE gamma-ray emission is difficult to explain only with synchrotron radiation, and several alternative processes have been proposed, such as synchrotron self-Compton (SSC), external inverse-Compton, proton synchrotron, and proton-induced cascade emissions~\citep[e.g.,][]{Nava2021,Gill2022}.
VHE gamma-ray observations will bring us new information on the physical mechanisms of GRBs, including both dynamics and particle acceleration and they can be used for testing fundamental physics.

GRB~221009A was an extremely energetic event.
With its redshift of 0.1505 \citep{Castro2022}, 
the  
isotropic-equivalent gamma-ray
energy is at least
$E_{\rm iso,\gamma} \approx 1.0-1.2\times 10^{55}$~erg \citep{Frederiks2023,Burns2023}.
The {\it Fermi} Large Area Telescope reported high-energy (HE: 0.1--10~GeV)
gamma-ray photons,
and the highest-energy photon reached around 100 GeV \citep{Pillera2022}. 
The Large High Altitude Air Shower Observatory (LHAASO) detected
more than 5000 VHE gamma-ray photons above 500 GeV within 2000~s 
after the GBM trigger, 
and the highest photon energy reached $\approx 18$~TeV \citep{Huang2022}.
The HE and VHE gamma-ray light curves may be affected by the $\gamma\gamma$ annihilation 
with source photons
\citep[][]{Murase2022,BTZhang2022} 
and the extragalactic background light \citep[EBL; e.g.,][]{Murase2007}.
These observations could also provide constraints on the model parameters such as the radiation region, the bulk Lorentz factor, and the jet opening angle.

Multiwavelength afterglow emission of GRB~221009A has been observed in HE gamma-ray, X-ray, optical, and radio bands
\citep[e.g.,][]{Fulton2023,Williams2023,Shrestha2023,Laskar2023,Kann2023,Levan2023}.
The X-ray and optical luminosities are
much brighter than those of typical long GRBs \citep{Ror2022}, 
while the radio (15 GHz) luminosity 
($\approx 1\times10^{29}~{\rm erg~s^{-1}~Hz^{-1}}$ at 10~days)
 is lower than typical long GRBs including VHE gamma-ray event
 [e.g.,
 $\approx 4\times10^{30}~{\rm erg~s^{-1}~Hz^{-1}}$ for 
  GRB~180720B 
\citep{Rhodes2020}].
Furthermore, around $4\times10^4$~s, the High-Altitude Water Cherenkov Observatory obtained 
the flux upper limit \citep{HAWC2022}, which is converted to the luminosity of $\approx 1\times10^{18}~{\rm erg~s^{-1}~Hz^{-1}}$ at 1 TeV.
This value is smaller than that of GRB~180720B ($\approx 3\times10^{20}~{\rm erg~s^{-1}~Hz^{-1}}$ at $3.6\times10^4$~s), 
and it is almost comparable to that of 
GRB~190829A ($\approx 3\times10^{17}~{\rm erg~s^{-1}~Hz^{-1}}$ at $2.7\times10^4$~s). 
These properties of the observed afterglow emission give us constraints on the modeling. 
A possible interpretation for GRB 221009A is that a
uniform
jet propagates into a wind medium \citep{Ren2022}.

In this Letter, to explain multiwavelength afterglow emission of GRB~221009A, we consider a two-component jet model, in which two top-hat jets with different opening angles propagating into uniform interstellar medium 
\citep[ISM; e.g.,][]{Ramirez-Ruiz2002,Berger2003,Huang2004,Peng2005,Wu2005,Racusin2008,Sato2021,Sato2022,Rhodes2022}.
GRB~221009A is among bursts with the largest 
$E_{\rm iso,\gamma}$
\citep[e.g.,][]{Atteia2017,Zhao2020}.
This suggests that the GRB jet has a small initial opening half-angle $\theta_0$ so that we get a normal value of the collimation-corrected gamma-ray energy, $E_{\gamma}=E_{{\rm iso},\gamma}\theta_0^2/2$ 
$\sim10^{50}$~erg \citep[see Figure 1 of][]{Zhao2020}. 
For example, if we take a typical value, $\theta_0\approx 0.1$~rad,
then we get an extremely large value, $E_{\gamma}\approx5\times10^{52}$~erg \citep[e.g.,][]{Zhao2020}.
However, due to the jet break effect, afterglow emission from a jet with a small opening angle decays more rapidly than observed. 
With another less collimated jet component, late-time afterglow emission can be as bright as observed \citep{Sato2021,Sato2022}.
Furthermore, we investigate the detectability of SSC photons with $\mathcal{O}$(10~TeV). 
In this Letter, cosmological parameters, $H_0 = 71~{\rm km}~{\rm s}^{-1}~{\rm Mpc}^{-1}$, $\Omega_M$ = 0.27 and $\Omega_{\Lambda}$ = 0.73 are adopted.



\section{Model description}
In this Letter, we consider the two-component jet model,
in which the overall flux
 is simply given by a superposition of emission from
 two relativistically moving top-hat jets.
Here, we shortly summarize 
the calculation of the emission from a single top-hat jet
\citep[see][for details]{Huang2000}.
Taking into account the radiative losses, 
we calculate the dynamics of the jet
(that is characterized only by the evolution of the shock radius, the bulk Lorentz factor, and the jet opening half-angle) and resulting multiwavelengths afterglow emission.
Unless otherwise stated, it is assumed that the jet propagates into the uniform surrounding material, that is, the ISM with constant density $n_0$.
The jet is assumed to have the isotropic-equivalent kinetic energy $E_{\rm{iso,K}}$, the bulk Lorentz factor $\Gamma_0$, and the opening half-angle $\theta_0$. 
We also assume a power-law electron energy distribution with a spectral index $p$, and constant microphysics parameters $\epsilon_e$, $\epsilon_B$, and $f_e$, which are the energy fractions of the internal energy going into non-thermal electrons, magnetic fields, and the number fraction of accelerated electrons, respectively.
Then, the synchrotron and SSC emissions are numerically computed taking into account the Klein-Nishina effect
\citep[e.g.,][]{Nakar2009, Murase2010, Murase2011, Wang2010, BTZhang2021, Jacovich2021,Sato2022}.
The EBL absorbs the VHE gamma-ray photons. We calculate the EBL absorption following \citet{Franceschini2008} by using the PYTHON package `EBL table'\footnote{https://pypi.org/project/ebltable/}.
The flux density $F_{\nu}$ is obtained by integrating the emissivity along the equal arrival time surface 
\citep[e.g.,][]{Granot1999}.
GRB~221009A was so bright that the viewing angle, which is the angle between the jet axis and the observer's line of sight, can be taken as $\theta_v=0$.


\section{Results of Afterglow Modeling}
In this section, we show our numerical results of multiwavelength afterglow emission in VHE gamma-ray (1~TeV), HE gamma-ray (1~GeV), X-ray (1~keV), 
optical ({\it r} band) and radio 
(1.3, 5.0, 15.8 and 99.9~GHz)
bands, and compare them with the observed data of GRB~221009A.
The X-ray data are taken from the {\it Swift} team website\footnote{https://www.swift.ac.uk/xrt\_curves/01126853/} \citep{Evans2007,Evans2009}.
We convert the observed energy flux in 0.3--10~keV to the flux density at 
1~keV, 
assuming that the photon index is 1.8  
during slow-cooling epochs.
The VHE gamma-ray upper limit is obtained from 
\citet{HAWC2022}.
Observed HE gamma-ray flux is taken from \citet{Ren2022}.
The observed energy flux in 0.1--10~GeV is converted to the flux density at 1~GeV with the photon index of 1.87 \citep{Pillera2022}.
Optical 
data are extracted from
\citet{Fulton2023} and \citet{Laskar2023}.
We adopt the r-band extinction 
$A_r = 4.31$~mag \citep{Ren2022}.
Radio 
data is taken from
\citet{Laskar2023}.


\subsection{One-component jet model}
\label{subsec:single}
\begin{figure*}
\vspace*{5pt}
\hspace*{-1cm}
\begin{minipage}{0.33\linewidth}
\centering
\includegraphics[width=1.0\textwidth]{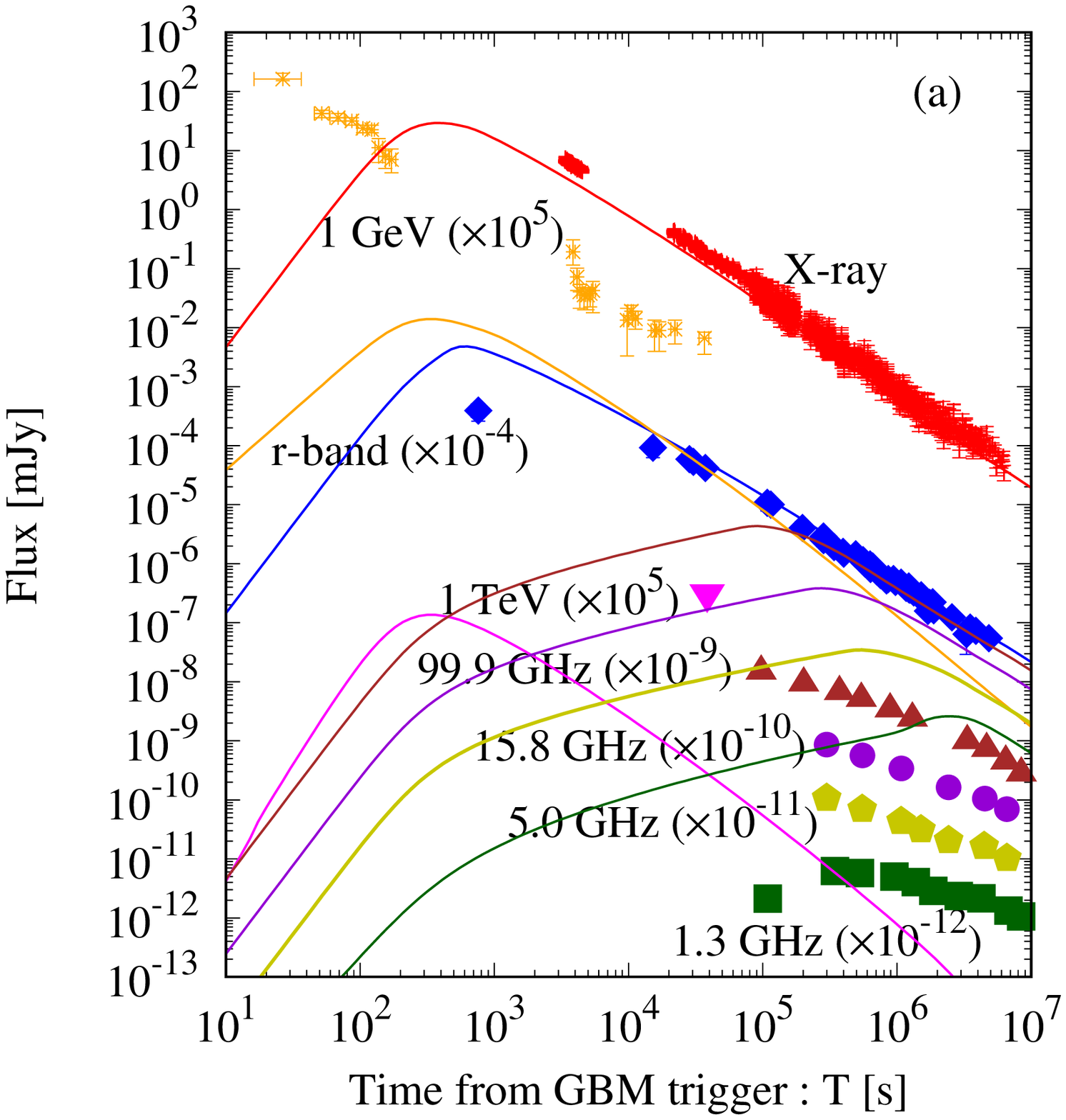}
\end{minipage}
\begin{minipage}{0.33\linewidth}
\centering
\includegraphics[width=1.0\textwidth]{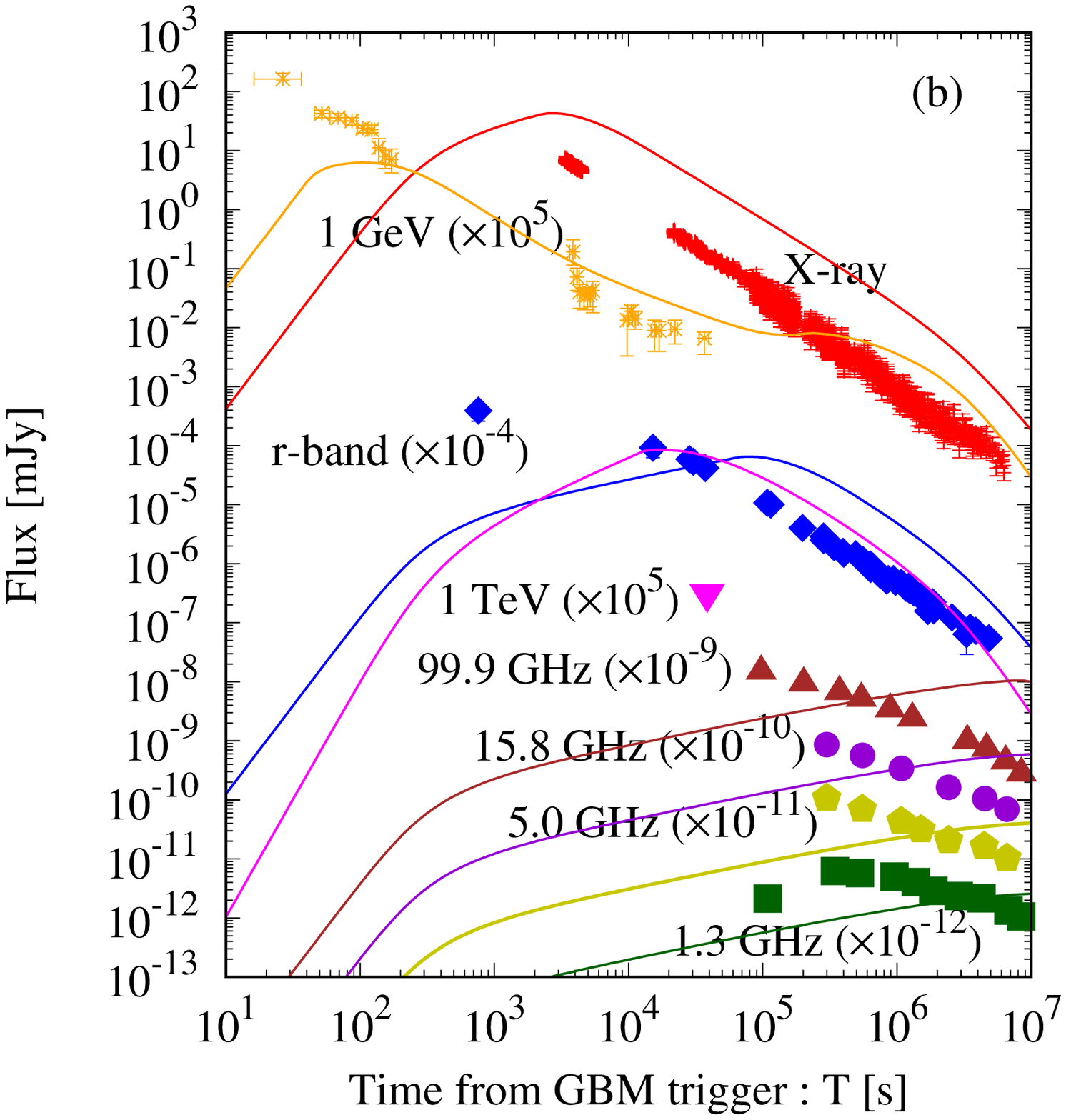}
\end{minipage}
\begin{minipage}{0.33\linewidth}
\centering
\includegraphics[width=1.0\textwidth]{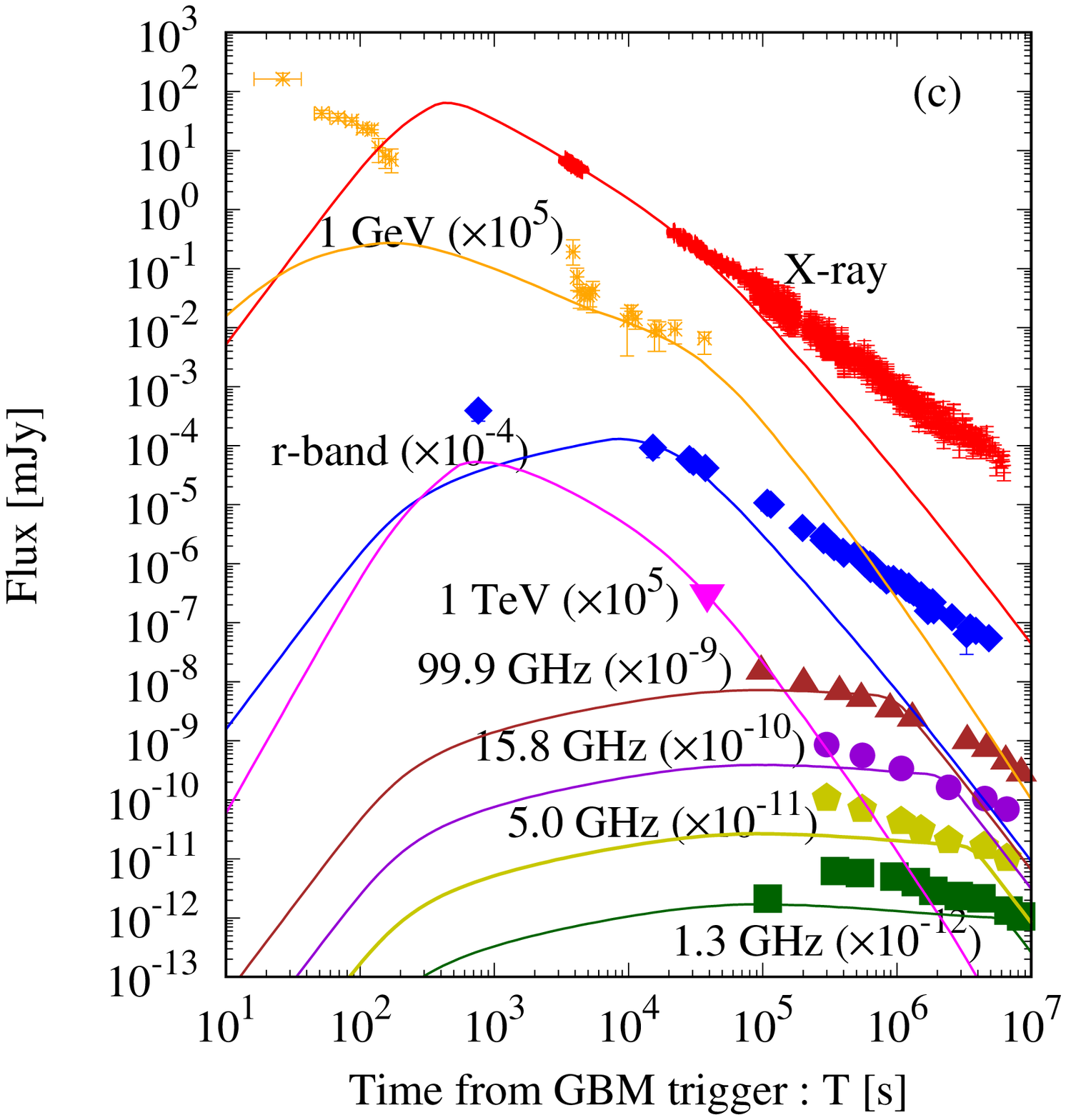}
\end{minipage}
\vspace{-0.3cm}
\caption{
Multiwavelength afterglow light curves of GRB~221009A
from models with 
single top-hat jets.
Theoretical results in VHE gamma-ray~(1 TeV: magenta), HE gamma-ray~(1 GeV: orange), X-ray~(1 keV: red), optical ({\it r} band: blue) and radio bands (1.3~GHz: green, 5.0~GHz: yellow, 
15.8~GHz: violet, and 
99.9~GHz: brown) are compared with the observations [1~TeV (upper limit): magenta downward triangle, 1~GeV: orange points, X-ray: red points, {\it r} band: blue diamonds, 
99.9~GHz: brown upward triangles, 
15.8~GHz: violet circles, 
5.0~GHz: yellow pentagons and 
1.3~GHz: green square]. The solid lines in panels~(a), (b) and (c) show the fluxes from the typical jet~I,
II, and the narrow jet, respectively.
All three models have difficulty in explaining the observed data.
}
\label{afterglow}
\end{figure*}

Before going to the two-component jet model, here
we discuss the emission from  the single top-hat jet model to show
 that the latter hardly explains the multiwavelength afterglow emission of GRB~221009A.
Given that the isotropic energy of the prompt emission $E_{{\rm iso},\gamma}\sim10^{55}$~erg, it is natural to set $E_{\rm iso,K}\sim10^{55}$~erg so that the efficiency of the prompt emission, $E_{{\rm iso}, \gamma}/(E_{\rm iso, \gamma}+E_{\rm iso, K})$, can be reasonable. Note that the isotropic-equivalent kinetic energy adopted by \citet{Ren2022} is smaller than ours, in which a very high radiative efficiency is indicated.  

First, we consider the jet with a typical initial opening half-angle, $\theta_0=0.1$~rad, propagating into a uniform ISM. 
In order to fit the observed X-ray and r-band light curves,
we set model parameters as
$\theta_v=0$, 
$E_{\rm iso,K}=1.0\times10^{55}$~erg, $\Gamma_0=285$, 
$n_0 =1.0\times10^{-2}$~cm$^{-3}$, $p=2.7$, 
$\epsilon_e=8.0\times10^{-4}$, $\epsilon_B=4.0\times10^{-3}$,
and $f_e=0.1$.
The jet with these parameters is referred to as `typical jet I' in the following.
Figure~\ref{afterglow}(a) shows the result.
While the  jet is in the adiabatic deceleration phase,
the cooling frequency $\nu_c$, the typical frequency $\nu_m$ and 
the absorption frequency $\nu_a$ obey the relation $\nu_a<\nu_m<\nu_c$.
The cooling frequency $\nu_c$ is 
between the {\it r} band and X-ray bands, and $\nu_m$ is lower than the {\it r} band.
The X-ray and optical light curves follow the scalings $F_\nu \propto T^{(2-3p)/4}\sim T^{-1.5}$ 
and $F_\nu\propto T^{3(1-p)/4}\sim T^{-1.3}$, respectively \citep[e.g.,][]{Gao2013},
which is 
roughly
consistent with the observed data.
However, the numerical results in 
1.3, 5.0, 15.8 and 99.9~GHz
bands 
are about 
three
orders of magnitude brighter than the observed data (green, yellow, violet, and brown lines in Fig.~\ref{afterglow}(a)).
Moreover, our numerical HE gamma-ray light curve is about 
an order of magnitude
dimmer than the observed HE gamma-ray flux (orange line in Fig.~\ref{afterglow}(a)).
Note that the value of $\epsilon_e/f_e=8.0\times10^{-3}$ is 
unusually small.

Secondly, we change microphysics parameters to fit the observed 
dim radio emission.
If we take larger $\epsilon_e$, smaller $\epsilon_B$,
and/or smaller $f_e$, then
the radio flux becomes small.
We adopt 
$\theta_v=0$, 
$\theta_0=0.1$~rad, 
$E_{\rm iso,K}=1.0\times10^{55}$~erg, $\Gamma_0=285$, 
$n_0 =1.0\times10^{-2}$~cm$^{-3}$, $p=2.7$, 
$\epsilon_e=0.4$,
$\epsilon_B=1.0\times10^{-6}$, and 
$f_e=0.08$
(we call this `typical jet II').
As shown in Fig.~\ref{afterglow}(b),
the radio emission can be dimmer.
However, our VHE gamma-ray, HE gamma-ray, X-ray and optical fluxes
overpredict the observations.
Moreover, our radio fluxes are still inconsistent with the observational results.
Therefore, models with  typical jets I and II having
$\theta_0=0.1$~rad are excluded.

Next, we try to explain radio afterglows with another parameter set.
Once $E_{\rm iso,K}$ is large, we need small
$n_0$ and/or $\theta_0$ to have dim radio emission.
The observed small radio fluxes require the jet with a small initial opening half-angle \citep{Sato2021,Sato2022}.
Here, we introduce `narrow jet' with $\theta_0=0.01$~rad,
and we set 
$\theta_v=0$, 
$E_{\rm iso,K}=1.0\times10^{55}$~erg, $\Gamma_0=285$, 
$n_0 =1.0\times10^{-2}$~cm$^{-3}$, 
$p=2.7$, $\epsilon_e=0.1$, 
$\epsilon_B=2.5\times10^{-5}$, and 
$f_e=0.2$
to fit the 
HE gamma-ray 
light curve.
It is found from Figure~\ref{afterglow}(c) that
our narrow jet is hard to explain the observed late ($T\ga3\times 10^4$~s) X-ray and optical afterglows.
The jet break time is given by $T_{\rm jet}\sim (3E_{\rm iso,K}/4\pi n_0m_pc^5)^{1/3}\theta_0^{8/3}$, where $m_p$ is the mass of the proton and $c$ is the speed of light \citep{Sari1999}.
For our narrow jet, 
we obtain $T_{\rm jet}\approx 2.4\times10^3$~s.
After the jet break, 
the X-ray and optical fluxes decay much steeper than the
observed ones (orange, red, and blue lines in Fig.~\ref{afterglow}(c)).
The initial opening half-angle of the narrow jet is rather small.
However, our narrow jet is fat in a sense, 
$\theta_0>\Gamma_0^{-1}$.

As shown earlier, the cases of constant ISM hardly explain the
observed afterglow emissions.
Another possibility is to consider a wind-like circumstellar medium
\citep{Ren2022}.
Then,  the late X-ray, optical, and radio afterglows are better
explained.
Even in this case, however, the predicted HE gamma-ray flux after $T\ga2\times10^4$~s
is 
about six times 
dimmer than the observational result.
We tried various parameter sets to fit the data,
however, we did not find better combinations than those described earlier
for the uniform density case. 
Hence, it is challenging for the
single top-hat jet model 
to well describe observed multiwavelength afterglow emission.
This conclusion is consistent with analysis done by \citet{Laskar2023}.
In our previous work, `two-component jet model', 
in which another wider jet is added to the narrower component,
was required to explain the observed afterglows of 
the other VHE gamma-ray events reported so far
\citep{Sato2021,Sato2022}.
In the next subsection, we will investigate 
whether GRB 221009A can be described by the same model.

\subsection{Two-component jet model}
\begin{figure}
\vspace*{10pt}
\hspace{-2cm}
\centering
\includegraphics[width=0.40\textwidth]{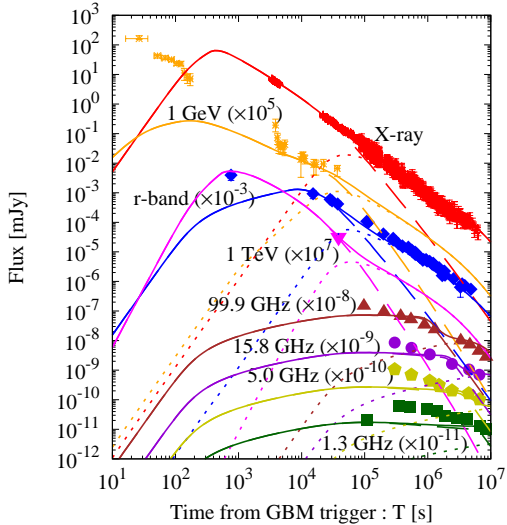}
\vspace{-0.7cm}
\caption{
Multiwavelength afterglow light curves calculated by the two-component jet model.
The meanings of colors and symbols are the same as in Fig.~\ref{afterglow}.
The solid lines are the sum of the narrow (dashed lines) and wide (dotted lines) jet components. 
}
\label{twojet}
\end{figure}

In this subsection, we show the result of our two-component jet model, in which another `wide jet' emission is introduced in addition to the narrow jet given in \S~3.1.
It is assumed that the two jets are on co-axis and ejected from the central engine simultaneously.
The observed flux can be described by a superposition of emission components from the two independent jets until $T\sim1\times10^7~{\rm s}$. 
This is because the solid angle of the narrow jet is small enough compared with that of the wide jet until $T\sim1\times10^7~{\rm s}$ in our model parameter set. 
After this time-scale, the expansion of the narrow jet affects the dynamics of the wide jet. For the narrow jet, we use the parameters determined in \S~3.1.
The parameters of the wide jet are 
$\theta_v=0$, 
$\theta_0=0.1$~rad, 
$E_{\rm iso,K}=2.0\times10^{53}$~erg, $\Gamma_0=24$, 
$n_0 =1.0\times10^{-2}$~cm$^{-3}$, 
$p=2.4$, $\epsilon_e=0.4$, $\epsilon_B=4.0\times10^{-5}$, and $f_e=0.1$.
The constant ISM density is considered in this subsection as well as in \S~3.1.
The microphysics parameters for the narrow and wide jets have different values,
although the differences are small.
If the circumburst or ejecta of the narrow and wide jets are different,
the two jets may not have common microphysics parameters \citep{Sato2022}.
Indeed, some authors have adopted different values
of $\epsilon_e$, $\epsilon_B$ and $f_e$ for the two jets
\citep[e.g.,][]{Racusin2008,Sato2021,Sato2022,Rhodes2022}.

As seen in Fig.~\ref{twojet}, the narrow jet emission is consistent with the early X-ray and optical 
data 
(dashed lines).
Moreover, the late 
X-ray and optical data points are explained by the wide jet emission (dotted lines).
At 
$T\approx 3\times10^4~{\rm s}$,
the wide jet enters the adiabatic expansion phase. 
Subsequently, $\nu_m$ intersects the {\it r}-band at 
$T\approx 5\times10^4~{\rm s}$. 
After that, $\nu_m$ is below the {\it r}-band and $\nu_c$ is between the optical and X-ray bands, so that the X-ray and {\it r}-band light curves follow 
$F_\nu\propto T^{(2-3p)/4}\sim T^{-1.5}$ and $F_\nu\propto T^{3(1-p)/4}\sim T^{-1.3}$,
respectively.
The late X-ray and {\it r} band data are well explained by the wide jet.
The electron spectral indices of the narrow jet 
($p=2.7$)
are roughly consistent with an observed X-ray photon index of 1.8.

Although the observed HE gamma-ray flux is much brighter than our afterglow model before $T\sim 10^3~{\rm s}$, such early HE gamma-ray emission may largely originate from inner jets and/or the external reverse shock \citep{Ren2022,BTZhang2022}. 
The reverse shock emission could also contribute to the observed flux in the optical band at $T \approx 3\times 10^3~{\rm s}$. 
The optical data may also be prompt optical emission.
The numerical radio fluxes are sometimes smaller than the observed ones
(brown, violet, yellow and green lines in Figs.~\ref{twojet} and 
A1 in the online material).
The radio emission may be another component \citep{Laskar2023,O'Connor2023}.

\begin{figure}
\vspace*{10pt}
\hspace{-2cm}
\centering
\includegraphics[width=0.4\textwidth]{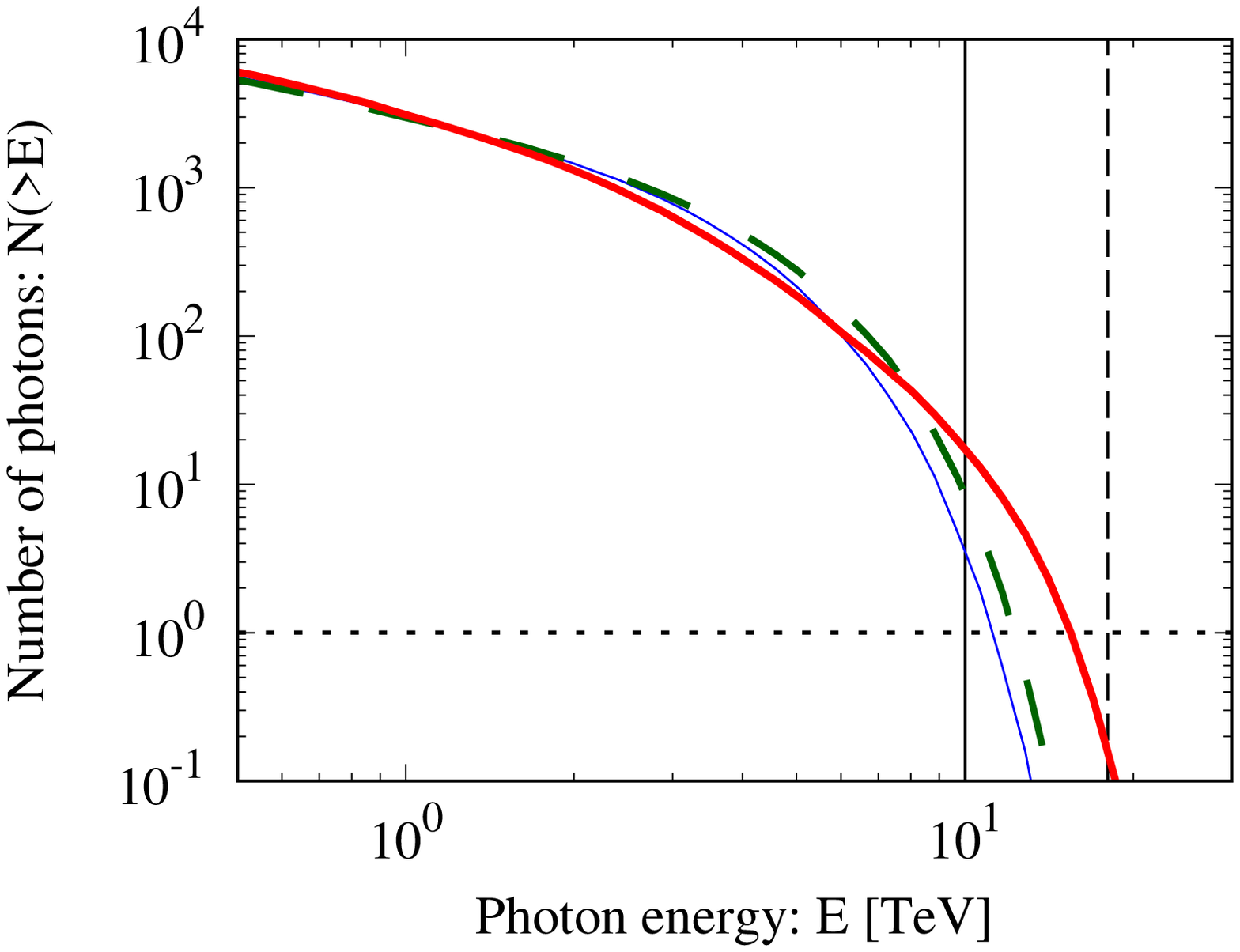}
\vspace{-0.5cm}
\caption{
The expected number of VHE photons detected by LHAASO for the narrow jet until 
2000~s 
after the {\it Fermi}/GBM trigger.
The thin-blue-solid \citep{Franceschini2008}, thick-green-dashed \citep{Gilmore2012}, and thick-red-solid \citep{Finke2010} lines are for different EBL models.
The vertical black solid and dashed lines
represent photon energies ($E$) of $E=10$~TeV and 18~TeV, 
respectively.
The horizontal dotted line shows $N(>E)=1$.
}
\label{number}
\end{figure}

\section{Summary and Discussion}
Using a two-component jet model, we have modeled multiwavelength afterglow of GRB~221009A assuming that the ambient matter is homogeneous. 
We have shown that 
the observed  HE gamma-ray, X-ray and optical afterglows are explained by the sum of the narrow and wide jet components.
For our model parameters, the collimation-corrected kinetic energy, $E_{\rm jet,K}= E_{\rm iso,K}\theta_0^2/2$, of the narrow jet is estimated to be $5.0\times10^{50}$~erg,
but it is still about five times larger than the normal value.
Our narrow jet may have large values of $E_{\rm iso,K}$, which are comparable to the observed isotropic-equivalent gamma-ray energy of the prompt emission, $E_{{\rm iso}, \gamma}$, which is among the largest. However, since its jet opening angle is small, the value of $E_{\rm jet,K}$ as well as the collimation-corrected gamma-ray energy of the prompt emission can remain normal, making energetics requirements reasonable.

We also evaluate the number of VHE gamma-ray photons, which can be detected by LHAASO.
In Fig.~\ref{number}, we show the result for 
2000~s 
after the {\it Fermi}/GBM trigger, for the narrow jet~\citep[e.g.,][]{BTZhang2022}. 
For the EBL absorption, three different models are adopted.
The effective areas of the LHAASO Water Cerenkov Detector Array for the zenith angle between 15 and 30~deg and that of the LHAASO larger air shower kilometer square area are obtained from \citet{Wang2022} and \citet{Ma2022}, respectively.
For our parameter set, the emission radius is so large that the 
intra-source
$\gamma \gamma$ annihilation
\citep{Svensson1987} would not strongly affect the VHE gamma-ray flux \citep[see Section~4.1 in][]{BTZhang2022}. 
Our results indicate that $\ga500$~GeV gamma-ray photons from the SSC component of the narrow jet could be detected by the LHAASO with a significance level $>100$~s.d. in the Li-Ma significance; otherwise the VHE observations can constrain our model. 

VHE gamma-ray and radio data 
could be
relevant for discriminating the two-component jet model from the other models. 
Radio emission from the narrow jet decays rapidly after $T\ga10^6$~s. 
The wide jet enhances the radio flux until $T\sim10^7$~s.
In this case, the 
transition from the narrow to wide jet components
may appear as shown in 
Figs.~\ref{twojet} and A1, 
which may be used 
as a test for
the existence of the wide jet by radio observations. 
However, the radio data may also come from another component, e.g., reverse shock \citep{O'Connor2023}, and numerical radio fluxes that are not compatible with the observed radio data reported by \citet{Laskar2023} do not mean that the model is excluded. 
The early ($T\la10^3$~s) VHE gamma-ray light curve shows rising both in the constant ISM and in the wind circumburst medium cases.
The slope is much shallower in the latter case than in the former \citep{Ren2022}. The temporal evolution in the rising part in VHE gamma-ray band may become diagnostic to discriminate between the two cases.

\citet{Sato2022} considered the two-component jet model for the other VHE GRBs that were reported previously. They found that all VHE gamma-ray events have similar values of $E_{\rm jet,K}$ for a wide jet component. For a narrow jet component, the collimation-corrected kinetic energy 
of GRB 221009A is about an order of magnitude larger than that of the other VHE gamma-ray events, which may suggest that GRB 221009A has the largest isotropic-equivalent gamma-ray energy of the prompt emission among the known VHE GRBs.
For all VHE gamma-ray events, the radio fluxes were observed, where they showed the difficulty for the standard afterglow model in explaining VHE gamma-ray and radio afterglows simultaneously. However, the two-component jet model could describe such complicated multiwavelength light curves from radio to VHE gamma-ray bands \citep{Sato2022}.
Our result indicates that VHE gamma-ray events may commonly consist of a structured jet that can be resembled by two jet components with different angular sizes and bulk Lorentz factors.

\section*{Acknowledgments}
We thank 
Katsuaki~Asano,
Shigeo~S.~Kimura,
and Shuta~J.~Tanaka for valuable comments.
We also thank the anonymous referee for his/her/their helpful comments to 
improve the paper. 
This research was partially supported by JSPS KAKENHI Grant 
Nos.~22J20105 (Y.S.), 20H01901 (K.M.), 20H05852 (K.M.), 
19H01893 (Y.O.), 
21H04487 (Y.O.), and 22H01251 (R.Y.).
The work of K.M. is supported by the NSF Grant 
Nos. AST-1908689, AST-2108466 and AST-2108467.

\section*{DATA AVAILABILITY}

The theoretical model data underlying this article will be shared on reasonable request to the corresponding author.


\appendix
\section{ONLINE MATERIAL}
Enlarged view of Fig.~2 in the radio bands  (1.3, 5.0, 15.8 and 99.9~GHz) is shown in Fig. A1.
\begin{figure}
\vspace*{10pt}
\hspace{-2cm}
\centering
\includegraphics[width=0.40\textwidth]{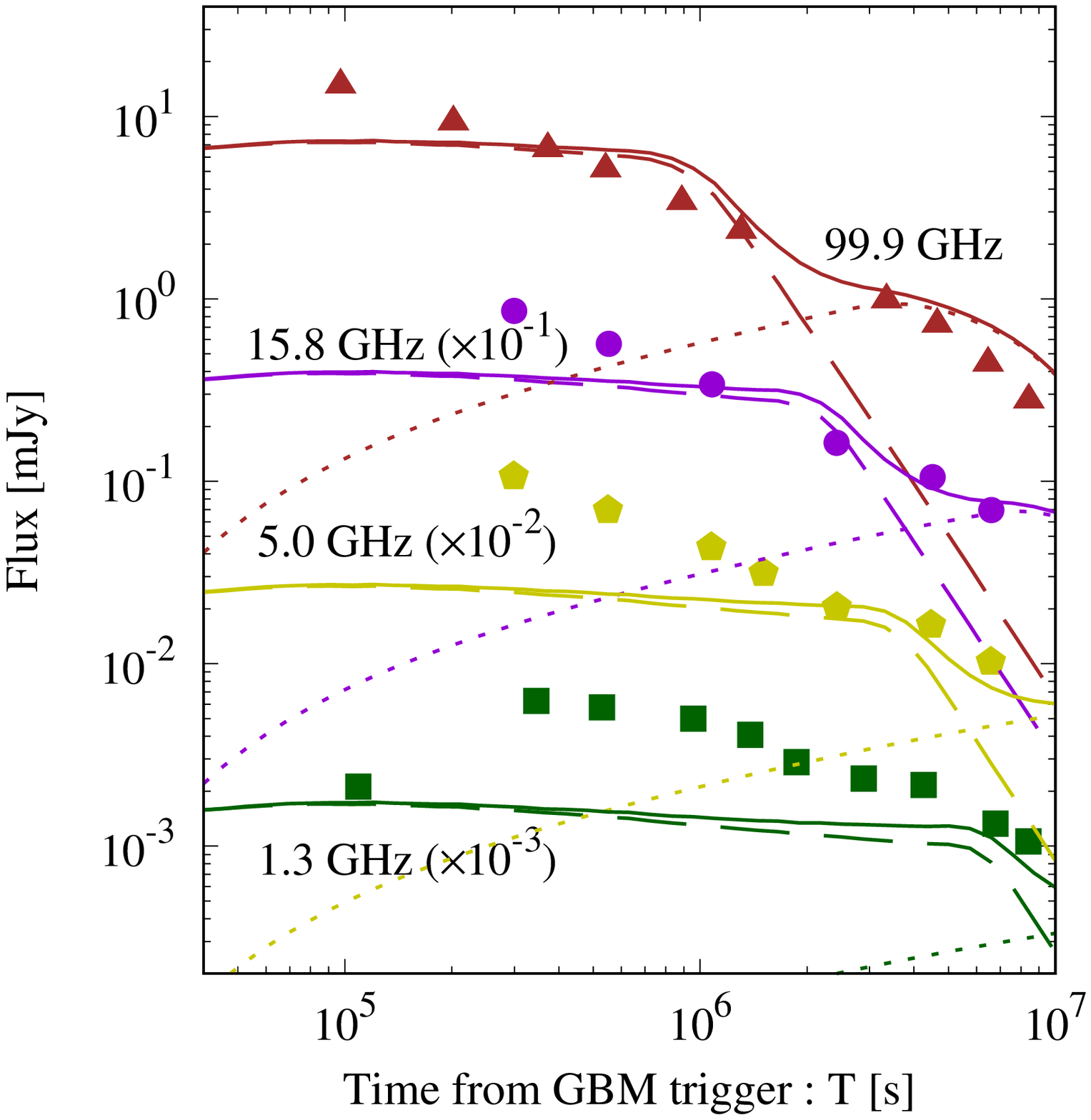}
\vspace{-0.7cm}
\caption{
Enlarged view of Fig.~2 in the radio bands  (1.3, 5.0, 15.8 and 99.9~GHz).
}
\label{twojetr}
\end{figure}

\label{lastpage}

\onecolumn

\twocolumn

\end{document}